\font\manual=manfnt
\def\dbend{{\manual\char127}}

\def\Z{\mathbf{Z}}

\def\r#1#2{``#1''$\rightarrow$``#2''}
\documentclass[conference]{IEEEtran}
\IEEEoverridecommandlockouts
\usepackage[hyphens]{url}
\usepackage{pdfpages}
\usepackage{verbatim}
\usepackage{hyperref}
\usepackage{combelow}
\usepackage[show]{ed}
\usepackage{graphicx}

\newtheorem{theorem}{Theorem}
\newtheorem{lemma}{Lemma}
\newtheorem{remark}{Remark}
\newtheorem{algorithm}{Algorithm}
\newtheorem{corollary}{Corollary}
\newtheorem{definition}{Definition}

\newtheorem{example}{Example}

\newtheorem{problem}{Problem}

\begin{document}
\author{\IEEEauthorblockN{
James H. Davenport}
\IEEEauthorblockA{\textit{Departments of Computer Science and Mathematical Sciences} \\
\textit{University of Bath}\\
Bath BA2 7AY, United Kingdom \\
masjhd@bath.ac.uk \quad  0000-0002-3982-7545}}
\title{Towards Verified Polynomial Factorisation\\
\thanks{The author is partially supported by EPSRC under grant EP/T015713/1. This research started at Dagstuhl workshop 23401 \cite{Davenportetal2024a}, and the author is grateful to Alex Best and Edgar Costa for their collaboration there. Further development was done at the Hausdorff Institute supported by  Deutsche Forschungsgemeinschaft (DFG, German Research Foundation) under Germany's Excellence Strategy – EXC-2047/1 – 390685813. Jeremy Avigad and Kevin Buzzard have given useful suggestions, and Mario Carneiro has been a constant source of advice and implementation.}}
\maketitle
\def\r{$\rightarrow$}\noindent
\def\prsubset{\subset}
\begin{abstract}
Computer algebra systems are really good at factoring polynomials, i.e. writing $f$ as a product of irreducible factors. It is relatively easy to verify that we have a factorisation, but verifying that these factors are irreducible is a much harder problem. This paper reports work-in-progress to do such verification in Lean.
\end{abstract}
\section{Introduction}
\cite{Davenport2023b} proposed the problem of \emph{proving} that a polynomial factorisation is correct.
The base case is polynomials in $\Z[x]$.
\begin{problem}[Factorisation]\label{prob:fact}
Given $f\in\Z[x]$, write $f=\prod f_i$ where the $f_i$ are \emph{irreducible} elements of $\Z[x]$.
\end{problem}
 Verifying that $f=\prod f_i$ is, at least relatively, easy. The hard part is verifying that the $f_i$ are \emph{irreducible}.  The author knows of no implementation of polynomial factorisation that produces any evidence, let alone a proof, of this: the user is expected to take the system's word for it.
We note that \cite[end of \S 6.3]{BallarinPaulson1999} noted, but did not answer, the irreducibility question.

\par
It is normal to state that ``We may as well assume $f$ is square-free (this would be a rather separate verification question)'', but an unconditional proof of factorisation (or even irreducibility) would need to include this step somehow. While we might implement (and prove correct) Euclid's algorithm in the theorem prover, it is easier to ask the algebra system to give us $\lambda,\mu$ such that $\lambda f+\mu f'=1$, and verify this in the theorem prover. A new algorithm for this verification is given in \cite[\S IV.C]{Davenport2024b}.
\par
However, in this paper we wil not concern ourselves with the problem of proving that a number $p$ is prime: while an interesting problem in general \cite{CaprottiOostdijk2001b}, the primes we use are small enough that any method will do.
\section{The Standard Algorithm for Factoring}
The basic algorithm goes back to \cite{Zassenhaus1969}: step M is a later addition \cite{Musser1975a}, and the  H' variants are also later.
\begin{enumerate}
\item Choose a prime $p$ (not dividing the leading coefficient of $f$) such that $f\pmod p$ is also square-free.
\item\label{step:p} Factor $f$ modulo $p$ as $\prod f_i^{(1)} \pmod p$ where the $f_i$ are  irreducible..
\item[M)]Take five $p$ and compare the factorisations.
\item If $f$ can be shown to be irreducible from modulo $p$ factorisations, return $f$.
\item Let $B$ be such that any factor of $f$ has coefficients less than $B$ in magnitude, and $n$ such that $p^n\ge 2B$.
\item Use Hensel's Lemma to lift the factorisation to $f=\prod f_i^{(n)} \pmod {p^n}$
\item[H)]\label{step:H} Starting with singletons and working up, take subsets of the $f_i^{(n)}$, multiply them together and check whether, regarded as polynomials over $\Z$ with coefficients in $[-B,B]$, they divide $f$ --- if they do, declare that they are irreducible factors of $f$.
\end{enumerate}
\cite{Davenportetal2024a,Davenport2023b} pointed out that, at least in principle, there is enough material generated in this algorithm to verify its correctness.  However, there are significant complications in the details, so we shall study a slightly easier problem.
\begin{problem}\label{P:simple}
Prove that $f$ is irreducible, using a computer algebra system to produce a certificate.
\end{problem}
\section{A proof of irreducibility}
It is clearly sufficient, even if not efficient, to proceed as follows.
\begin{enumerate}
\item  Ask the algebra system for a factorisation 
	\begin{equation}\label{eq:1}
		f=\prod_{i=1}^k f_i.
\end{equation}
\item If $k>1$ verify that this is a factorisation, i.e that (\ref{eq:1}) is true.
\item  For $i=1\ldots k$ do:
\item[3.1)]
Ask the algebra system for hints, essentially a certificate that $f_i$ is irreducible;
\item[3.2)]
Verify these hints.
\end{enumerate}
The inefficiency  comes from the fact that step 3.1 is recomputing things, or variants of things, that were computed in step 1.
\par
So what might such a certificate be? We will call the polynomial to be certified $f$, rather than $f_i$ for some $i$.
\subsection{The Simple Certificate}\label{C:Simple}
This consists of a prime $p$, and the assertion that $f$ is irreducible modulo $p$. This is wonderful if it works, but there two obstacles. 
The first is that we may not find such a $p$ easily: if the Galois group of $f$ is the symmetric group $S_n$, the probability of $f$ being irreducible modulo a prime $p$ is $1/n$.  The second is that such $p$ may not even exist: \cite{SwinnertonDyer1969} shows how to construct $f$ with no such $p$.
\subsection{The pre-Musser Certificate}\label{C:preM}
This consists of a prime $p$, a number $n$, and a set of polynomials $f_j\in\Z[x]$ together with the following assertions.
\renewcommand{\labelenumi}{P.\arabic{enumi}}
\begin{enumerate}
\item $f=\prod f_j\pmod{p^n}$.
\item Each $f_j$, considered as a polynomial modulo $p$, is irreducible.\label{pMI}
\item Any factor of $f$ over $\Z$ must have coefficients $<p^n/2$ in absolute value.\label{Plast}
\item No proper (nontrivial) subset $\{f_k\}\prsubset\{f_j\}$ has the property that $\prod_kf_k$, considered as a polynomial in $\Z[x]$ with coefficients  $<p^n/2$ in absolute value, is a factor of $f$.
\end{enumerate}
\subsection{The Simple post-Musser Certificate}\label{C:M}
This consists of a number $k$, a set of primes $p_i: i=1\ldots k$, and some sets of polynomials $\{f_{i,j}: j=1\ldots n_i\} i=1\ldots k$  together with the following assertions. 
\renewcommand{\labelenumi}{S.\arabic{enumi}}
\begin{enumerate}
\item For every $i$, $f=\prod f_{i,j}\pmod{p_i}$.
\item Each $f_{i,j}$, considered as a polynomial modulo $p_i$, is irreducible.\label{pMII}
\item For each $k: 0<k<\deg f$ there is an $i$ such that the factorisation $f=\prod f_{i,j}\pmod{p_i}$ is incompatible with a factorisation of $f$ as a degree $k$ polynomial and a $\deg f -k$ co-factor.\label{SD}
\end{enumerate}
The classic example of a Simple post-Musser Certificate is when a degree 4 polynomial factors modulo $p_1$ as two irreducible quadratics and module $p$ as a linear times an irreducible cubic. Then the two quadratics rule out $k=1,3$ and $p_2$ rules out $k=2$.
\par
But the Swinnerton-Dyer polynomials are examples where condition \ref{SD} may never be met. The simplest example is $x^4+1$ which is irreducible, but factors as two quadratics, or more, modulo every prime. The standard process in computer algebra, if five \cite{Musser1975a} or seven \cite{LuczakPyber1997} primes don't give us such a certificate, is to take the prime with fewest factors, and develop a pre-Musser certificate with that.
\subsection{A Complex post-Musser Certificate}\label{C:Complex}
While initially researching this project, using FLINT \cite{FLINT2023a} as our computer algebra system, \cite{Davenportetal2024a} discovered that this can generate a more complex proof of irreducibility. It consists of the union of the data of the two previous certificates, and assertions P.1--P.\ref{Plast}, S.1--S.\ref{pMII}, and the  following merger of P.4 and S.3.
\renewcommand{\labelenumi}{C.\arabic{enumi}}
\begin{enumerate}
\item For each $k: 0<k<\deg f$ for which there isn't $i$ such that the factorisation $f=\prod f_{i,j}\pmod{p_i}$ is incompatible with a factorisation of $f$ as a degree $k$ polynomial and a $\deg f -k$ co-factor, all subsets $\{f_\ell\}\prsubset\{f_j\}$ such that $k=\sum\deg f_\ell$, have the property that $\prod_\ell f_\ell$, considered as a polynomial in $\Z[x]$ with coefficients  $<p^n/2$ in absolute value, is not a factor of $f$.
\end{enumerate}
In the case of a Swinnerton-Dyer polynomial, clause S.3 doesn't buy us anything (as stated there), and we revert to P.4. But there are polynomials in the FLINT test suite for which S.3 gives a significant improvement.
\section{Irreducibility Modulo $p$}
Both the basic certificate and step S.\ref{pMI} of the pre-Musser certificate (and similar steps in the post-Musser certificates) require proving that a polynomial $f$ is irreducible modulo $p$. 
\subsection{Distinct Degree Factorisation}
\begin{theorem}\label{Thm:DD}
For $i \ge 1$ the polynomial
\begin{equation}
{\displaystyle x^{q^{i}}-x\in \mathbf {F} _{q}[x]}
\end{equation}
is the product of all monic irreducible polynomials in $\mathbf {F} _{q}[x]$ whose degree divides $i$.
\end{theorem}
Then the usual algorithm for polynomial factorisation is this.
\begin{algorithm}[Distinct Degree Factorisation: \cite{CantorZassenhaus1981}]
Let $f(t)$ be a square-free polynomial.
We define $f_1(t) = f(t)$
and inductively for $j = 1, 2, 3, \ldots$ define 
\begin{equation}\label{CZ1}
u_j(t) = \gcd(f_j(t), t^{p^j} - t)\hbox{ and }f_{j+ 1}(t) =f_j(t)/ u_j(t).
\end{equation}
The iteration stops when $f_{j+1 }(t)$ is constant.
\par
Then $u_j(t)$ is the product of all the factors of $f$ of degree $j$.
\end{algorithm}
\begin{remark}\label{R1}
We shouldn't compute (\ref{CZ1}) directly, but rather via this reformulation:
\begin{equation}\label{CZ2}
u_j(t) = \gcd(f_j(t), t^{p^j} - t\pmod{f_j(t)}) 
\end{equation}
where $t^{p^j}$ is computed by repeated squaring \emph{and} reducing modulo $f_j(t)$ after each multplication. This applies also to the next result.
\end{remark}
\begin{corollary}[To Theorem \ref{Thm:DD}]
A polynomial $f$ of degree $n$ is irreducible modulo $p$ iff  $\gcd\left(f(t), t^{p^j} - t\pmod{f(t)}\right)=1$ for $1\le j \le \lfloor\frac n2\rfloor$.
\end{corollary}
Note that we have managed to drop the ``square-free'' hypothesis.
However, we still need to handle Remark \ref{R1}. Jeremy Avigad suggested that the best method might be to write a Lean program to compute  $ t^{p^j} - t\pmod{f(t)}$ and prove this is correct. Then we can verify coprimeness, either directly in Lean, or by asking computer algebra to tell us $\lambda, \mu$ such that $\lambda f(t)+\mu \left(t^{p^j} - t\pmod{f(t)}\right)=1$, and verify this.
\par
This is currently being developed, but requires a computational implementation of polynomials in Lean.
\section{Bounds}
Though Simple or Simple post-Musser Certificates \emph{may} suffice, the Swinnerton-Dyer polynomials show that we may need to rule out factorisations based on assertion P.4 (or its development in C.1),and this requires assertion P.3, which means computing $B$ bounding the coefficients of any factor of $f$.
This is ``well known'' in computer algebra, and much ingenuity goes into computing better versions of the bounds and/or special cases, since the computing time of factorisation is, in practice, $O(\log^2b)$.
\subsection{Formal Proof and ``Similarly''}
\begin{figure*}[t]
\caption{Mignotte Lemma 1a in Lean \label{MignotteL1a}}
\includegraphics{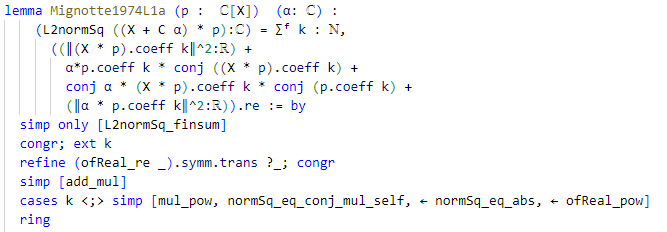}
\end{figure*}
\par
\cite{Mignotte1974} is a key piece of the bounds computation. It uses $||\cdot||$ for the $L_2$ norm of a polynomial.
\begin{lemma}
Let $P(X)$ be a polynomial with complex coefficients and $\alpha$ be a
nonzero complex number. Then
$$
||(X + \alpha)P(X)|| = |\alpha| ||(X + \overline\alpha^{-1})P(X)||.
$$
\end{lemma}
The proof goes as follows. \cite{Mignotte1974}  writes
\begin{eqnarray*}
P(X)&=&\sum_{k=0}^m a_kX^k,\cr
Q(X)&=&(X+\alpha)P(X)=\sum_{k=0}^{m+1}(a_{k-1}+\alpha a_k)X^k\cr
R(X)&=&(X + \overline{\alpha}^{-1})P(x)=\sum_{k=0}^{m+1}(a_{k-1}+\overline{\alpha}^{-1}a_k)X^k
\end{eqnarray*}
with the notation $a_{-1}=a_{m+1}=0$.
Then
$$
||Q||^2=\sum_{k=0}^{m+1}|a_{k-1}+\alpha a_k|^2=\sum_{k=0}^{m+1}(a_{k-1}+\alpha a_k)\overline{(a_{k-1}+\alpha a_k)}
$$
which expands to
\begin{equation}\label{L1a}
\sum_{k=0}^{m+1}\left(|a_{k-1}|^2+\alpha a_k\overline{a_{k-1}}+\overline\alpha a_{k-1}\overline{a_k}+|\alpha^2||a_k|^2
\right).
\end{equation}
This is accomplished in Lean by the code in Figure \ref{MignotteL1a}.
\par
\cite{Mignotte1974}  then says ``Expanding $|a |^2 ||R||^2$ yields the same sum''.
\par
However, if we expand $|a |^2 ||R||^2$ naively as above, we actually get 
\begin{equation}\label{L1b}
\sum_{k=0}^{m+1}\left(|\alpha^2||a_{k-1}|^2+\alpha a_k\overline{a_{k-1}}+\overline\alpha a_{k-1}\overline{a_k}+|a_k|^2
\right).
\end{equation}
In general (\ref{L1a}) and  (\ref{L1b}) are different: the $|\alpha|^2$ multiplies different terms. And indeed, for any $k$ the index-$k$ summands in  (\ref{L1a}) and  (\ref{L1b}) do differ.
However, it is legitimate to re-express  (\ref{L1a}) as  (\ref{L1a-bis}):
$$
\sum_{k=0}^{m+1}\left(|a_{k-1}|^2\right)+\sum_{k=0}^{m+1}\left(\alpha a_k\overline{a_{k-1}}+\overline\alpha a_{k-1}\overline{a_k}\right)+\sum_{k=0}^{m+1}\left(|\alpha^2||a_k|^2\right)
$$
and then as 
\begin{equation}\label{L1a-bis}
||P||^2+\sum_{k=0}^{m+1}\left(\alpha a_k\overline{a_{k-1}}+\overline\alpha a_{k-1}\overline{a_k}\right)+|\alpha^2|||P||^2.
\end{equation}
A similar operation on (\ref{L1b})  gives (\ref{L1b-bis}) :
\begin{equation}\label{L1b-bis}
|\alpha^2|||P||^2+\sum_{k=0}^{m+1}\left(\alpha a_k\overline{a_{k-1}}+\overline\alpha a_{k-1}\overline{a_k}\right)+||P||^2,
\end{equation}
and now the equality between (\ref{L1a-bis}) and (\ref{L1b-bis}) is obvious,

\subsection{Formal proof of the rest of the bound}
This is work in progress. Note that, for Problem \ref{P:simple}, we just need a reasonable bound, which could be computed by the formal prover (but at that stage the prover doesn't know whether it's needed: we may get a \ref{C:Simple} or \ref{C:M} certificate), or the algebra system could use the same bound as the formal prover.
\section{Conclusions}
\bibliography{../../../jhd}
\end{document}